\documentclass[article,12pt]{elsarticle}

\RequirePackage{amsthm,amsfonts,amssymb}
\RequirePackage[colorlinks,citecolor=blue,urlcolor=blue]{hyperref}
\RequirePackage{graphicx}

\usepackage[reqno]{amsmath}

\usepackage{enumerate}
\usepackage{dsfont}
\usepackage[usenames,dvipsnames]{color}
\usepackage{mathrsfs}
\usepackage{placeins}
\usepackage{graphicx}
\usepackage{tikz}
\usetikzlibrary{arrows}
\usepackage{url}
\usepackage{multicol}
\usepackage{cleveref}
\usepackage{stmaryrd}
\usepackage{makecell}
\usepackage{natbib}

\numberwithin{equation}{section}

\theoremstyle{plain} 
\newtheorem{theorem}{Theorem}[section]
\newtheorem{lemma}[theorem]{Lemma}
\newtheorem{proposition}[theorem]{Proposition}
\newtheorem{corollary}[theorem]{Corollary}

\theoremstyle{remark}
\newtheorem{definition}[theorem]{Definition}

\newtheorem{example}[theorem]{Example}

\newtheorem{remark}[theorem]{Remark}

\newcommand{\bthe}{\begin{theorem}}
\newcommand{\ethe}{\end{theorem}}

\newcommand{\ben}{\begin{enumerate}}
\newcommand{\een}{\end{enumerate}}

\newcommand{\bit}{\begin{itemize}}
\newcommand{\eit}{\end{itemize}}

\newcommand{\beq}{\begin{equation}}
\newcommand{\eeq}{\end{equation}}

\newcommand{\ble}{\begin{lemma}}
\newcommand{\ele}{\end{lemma}}

\newcommand{\bde}{\begin{definition}\rm}
\newcommand{\ede}{\halmos\end{definition}}

\newcommand{\bco}{\begin{corollary}}
\newcommand{\eco}{\end{corollary}}

\newcommand{\bpr}{\begin{proposition}}
\newcommand{\epr}{\end{proposition}}

\newcommand{\brem}{\begin{remark}\rm}
\newcommand{\erem}{\end{remark}}

\newcommand{\bproof}{\begin{proof}}
\newcommand{\eproof}{\end{proof}}

\newcommand{\bexam}{\begin{example}\rm}
\newcommand{\eexam}{\end{example}}

\newcommand{\bfi}{\begin{fig}}
\newcommand{\efi}{\end{fig}}

\newcommand{\btab}{\begin{tab}}
\newcommand{\etab}{\end{tab}}

\newcommand{\beao}{\begin{eqnarray*}}
\newcommand{\eeao}{\end{eqnarray*}\noindent}

\newcommand{\balo}{\begin{align*}}
\newcommand{\ealo}{\end{align*}}

\newcommand{\balm}{\begin{align}}
\newcommand{\ealm}{\end{align}\noindent}

\newcommand{\beam}{\begin{eqnarray}}
\newcommand{\eeam}{\end{eqnarray}\noindent}

\newcommand{\barr}{\begin{array}}
\newcommand{\earr}{\end{array}}

\newcommand{\R}{\mathbb{R}}

\def\cB{\mathcal{B}}

\def\bO{\boldsymbol O}

\def\bX{\boldsymbol X}
\def\bY{\boldsymbol Y}

\newcommand{\ytoinf}{\lim_{y\to+\infty}}

\newcommand{\vague}{\stackrel{\lower0.2ex\hbox{$\scriptscriptstyle
                    \it{v} $}}{\rightarrow}}
\newcommand{\weak}{\stackrel{\lower0.2ex\hbox{$\scriptscriptstyle
                    \it{w} $}}{\rightarrow}}
\newcommand{\what}{\stackrel{\lower0.2ex\hbox{$\scriptscriptstyle
                    \it{\hat{w}} $}}{\rightarrow}}
\newcommand{\eqdis}{\stackrel{\lower0.2ex\hbox{$\scriptscriptstyle
                    \mathrm{d}$}}{=}}
\newcommand{\distr}{\stackrel{\lower0.2ex\hbox{$\scriptscriptstyle
                    \it{d} $}}{\rightarrow}}
\allowdisplaybreaks 
\definecolor{darkgreen}{RGB}{0,139,0}

\begin{document}
\begin{frontmatter}

\title{A parsimonious tail compliant multiscale statistical model for aggregated rainfall}

\author[label1]{Pierre Ailliot} 
\affiliation[label1]{
organization={Univ Brest, CNRS UMR 6205, Laboratoire de Mathématiques de Bretagne Atlantique},
            country={France}}
            
\author[label2]{Carlo Gaetan} 
\affiliation[label2]{organization={Dipartimento di Scienze Ambientali, Informatica e Statistica, Universita Ca' Foscari di Venezia, Venice },
            country={Italy}}

\author[label3]{Philippe Naveau} 
\affiliation[label3]{organization={Laboratoire des Sciences du Climat et l'Environnement},
            country={France}}            

\begin{abstract}
Modeling the probability distribution of rainfall intensities  at different aggregation scales, say from sub-hourly to weekly,  has always played a key role in most hydrological  risk analysis, in particular in the computation of Intensity-Duration-Frequency (IDF)  curves.  
Since any aggregation procedure involves accumulating rainfall over a prescribed time window, it naturally induces simple mathematical constraints related to summation. In particular, return levels inferred from a statistical model should be ordered across time scales, reflecting for example  the fact that observed daily accumulations necessarily exceed those at sub-daily scales. From a statistical modeling perspective, each aggregation step combines information from shorter time scales without introducing additional data. Consequently, the number of model parameters should remain limited.
Still, parsimonious aggregation models that describe the full distribution of rainfall intensities are sparse in the hydrological literature. 
In particular, most studies focus on extremes, e.g. by taking seasonal block maxima at different aggregation scales. 

In this study, we propose a statistical framework that allows to 
 model all rainfall intensities (low, medium and large) at different aggregation scales, while being parsimonious.  
To reach this goal, we use the extended generalized Pareto distribution (EGPD), which complies with extreme value theory for both low and high extremes and is flexible enough to capture the bulk of the distribution. 
We show a general result that explains how EGPD random variables behave under different types of aggregation procedures. 
Direct likelihood inference is difficult in our setting. However, by linking the EGPD class to Poisson compound sums, we can use the Panjer algorithm to quickly and efficiently evaluate the composite likelihood of our proposed model. 
As a result, return levels can be obtained for any return period, particularly those below the annual and seasonal scales. In addition, 
our approach insures that return levels do not cross with aggregation. 

To demonstrate the applicability of our method, we analyze sub-hourly time series from six gauging stations in France that have different climatological features. For each station, we only need a total of eight parameters to capture aggregation scales from six minutes to three days.  
 IDF curves above and below the annual scale are provided. 
 
\end{abstract}


\end{frontmatter}

\section{Introduction}
Rainfall aggregation over different time scales is a ubiquitous topic in hydrology. 
Describing  changes in precipitation  distributions from, say sub-hourly to higher scales such as daily or weekly periods, represents an important statistical task in any hydrological risk analysis. 
The archetypal example of this practical and scientific endeavor is the extensive literature dedicated to the generation of Intensity-Duration-Frequency (IDF) curves.
These curves are fundamental tools in hydrology and water resources engineering, and illustrate the relationship between rainfall intensity, duration, and frequency (or return period).
Historically, the geophysical analysis of  IDF curves can be rooted back, at least, to   more than half century ago.
For example, \cite{usweather_tp40}  provided nationwide IDF curves for durations from 30 minutes to 24 hours and return periods up to 100 years.
Today,  one can find various flavors of IDF curves \cite[see, e.g.][]{haruna2023modeling},  their associated maps for different regions of the world, and also well documented   software packages for IDF computations \cite[see, e.g.][]{Ulrich20}.

A common fundamental thread in all IDF curve studies  is to determine what is the appropriate probability distribution   of precipitation sums (equivalently averages), and how the features of such distributions vary when  the chosen aggregation period varies, say from sub-hourly to weekly scales, see \cite{koutsoyiannis2024stochastics}.  
In this context, 
one main motivation of this work is to pinpoint a statistical paradox within extremes of  rainfall aggregates, and to propose a model that reconciles contradictory interpretations of high return levels. 
To explain such a paradox, we need to introduce the following notation. 
The capital letter $Y$ corresponds to the random variable of interest at the smallest time scale of interest, say sub-hourly precipitation. 
Then, the simplest way to make rainfall aggregates is to sum these sub-hourly  observations over a given period of interest of length $d$, by computing 
$Y_1+...+Y_d$.

By construction, we always have 
$
    Y_i  \leq Y_1+...+Y_d, 
$ for any $i=1,\dots,d$.
This constraint simply means that the sum of positive terms is always greater than any elements composing this sum\footnote{Although simple, this statement is very robust. Even if the non-negative sample $(Y_1,\dots,Y_d)$ corresponds to non-stationary data that may be strongly dependent, the inequality remains true as all $Y_i\geq0$.}. In terms of return levels and survival functions, this implies that for any $u\geq 0$
\begin{equation}\label{eq: constraint}
\Pr(Y_i >u)  \leq \Pr(Y_1+...+Y_d >u).
\end{equation}
Therefore,  the return levels\footnote{The return level  with respect to the return period $T$ corresponds to the scalar $u_T$ satisfying the equation $\Pr(Y>u_T)=1/T$, i.e. it is the $1-1/T$ quantile of the random variable $Y$
\cite[see, e.g.][]{katz:parlange:naveau:2002}. }
of $Y_i$ should never intersect the return levels of the sum.
It follows that any statistically coherent rainfall aggregation study should impose constraints \eqref{eq: constraint} for all values of $u$, even when $u$ is large. 
However, this is not always the case, as we will briefly illustrate. 

To estimate high return levels, most hydrological analyses rely on extreme value theory (EVT)  \cite[see, e.g.][]{coles-01,beirlant:goegebeur:teugels:segers:2004}. In particular, exceedances above a high threshold are classically modeled by 
the generalized Pareto (GP) cumulative distribution function   defined by 
\begin{equation}\label{eq: GP}
    {H}_{\xi}(y/\sigma)=1-\left(1+\xi\frac{y}{\sigma}\right)_+^{-1/\xi},   \quad \mbox{$\sigma>0$, $\xi\in\R$},
\end{equation}
with the notation $a_+=\max(a,0)$. The choice $\xi=0$ corresponds to the exponential case. 
The scale parameter $\sigma$ drives the spread among  extreme exceedances, while 
the shape parameter $\xi$ controls the upper tail behavior. Estimates of
 $\xi$   are often  positive for precipitation \cite[see, e.g.][]{katz:parlange:naveau:2002} and we assume that $\xi>0$ in this work. 

In the context of IDF studies, it is natural to ask how the shape parameter $\xi$ of the sum changes with the aggregation scale. In particular, is the shape parameter of the sum different from that of its components $Y_i$? 

\begin{figure}[t]
    \centering
    \includegraphics[scale=0.75]{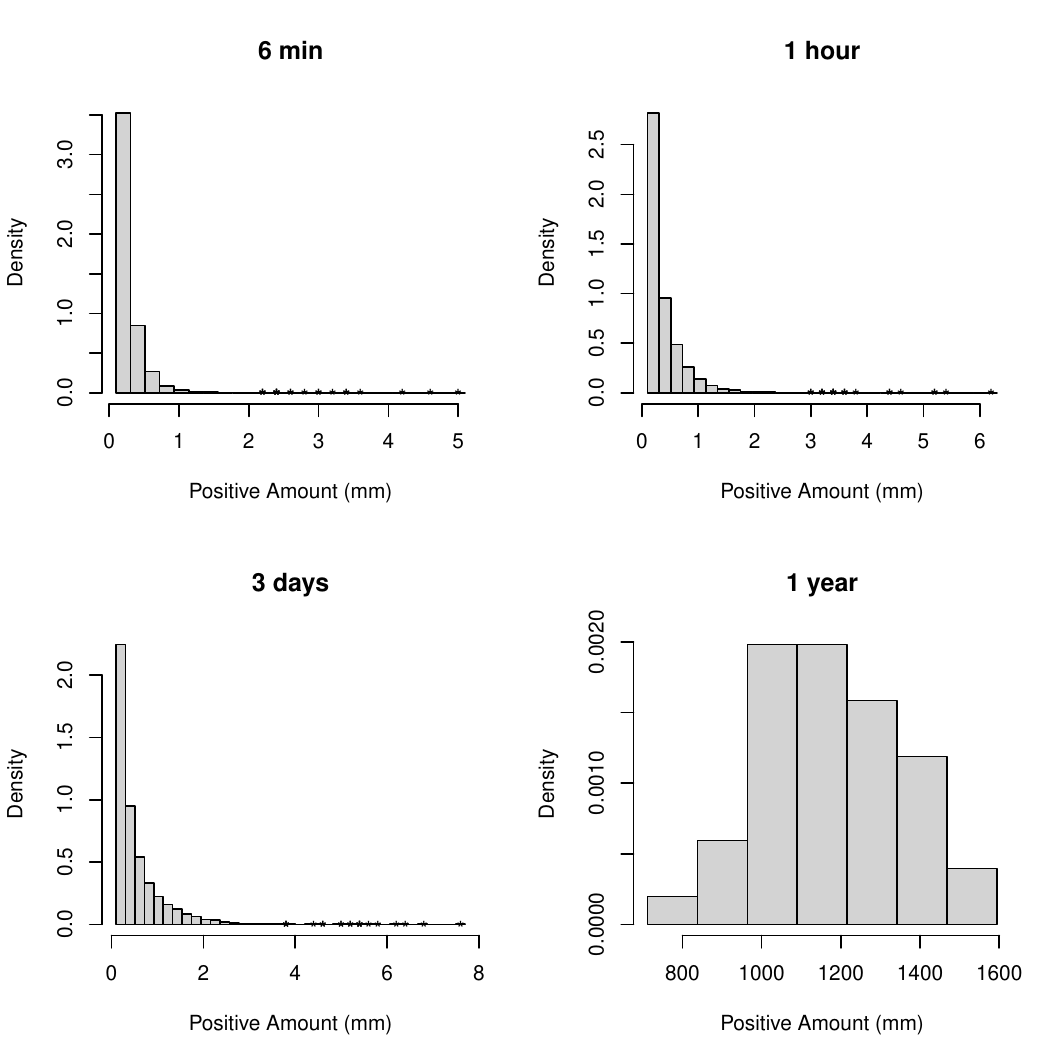}
    \caption{Histogram of positive rainfall (i.e. after removing dry) aggregated over different time scales in Brest. The stars on the x-axis correspond to the $20$ more extremes observations. The yearly histogram is plotted for illustration purpose, the modeling of such aggregation scale is not discussed in the paper. The results at the annual scale were obtained using $80$ years of daily data downloaded from \url{https://www.ecad.eu/}, whereas the other histograms are based on 18 years of 6-minute data for August-September described in Section~\ref{sec:data}.
    }
    \label{fig:expdf}
\end{figure}

To start answering  this question, one can look at a simple case study and compare the empirical distributions obtained at different aggregation scales.
For example, Figure \ref{fig:expdf} displays four  histograms of  rainfall rates recorded in Brest (France). The upper left panel begins with the aggregation scale of six minutes, while yearly aggregation scale is displayed in the bottom right panel. As anticipated from \eqref{eq: constraint}, the x-axis range (rainfall sums) increases with the increasing scale, but 
the yearly histogram appears closer to  a bell shape curve (normal distribution). 
This suggests that the upper tail behaviors of precipitation are affected by increasing scales. 

This phenomenon is well known by hydrologists working on extreme rainfall analysis. For example, \cite{haruna2023modeling} carefully modeled rainfall measurements from 81 weather stations in Switzerland with a minimum record length of 20 years. Figure 6 in that paper clearly indicates that the shape parameter $\xi$ decreases from 0.3 (30-minute scale) to essentially zero (daily scale).
Statistically, one could invoke the central limit theorem to explain this phenomenon. Sums or averages of random variables with finite variances should become closer to a Gaussian distribution as the aggregation scale increases.
 As the Gaussian distribution has a lighter tail than a Pareto tail \cite[see][for instance]{nair2022}, it seems to make sense  that  cumulative rainfall extremes appear to have a lighter tail than the ones at finer timescale. 
However, this reasoning contradicts  inequality \eqref{eq: constraint}!
\footnote{To see this,  EVT with \eqref{eq: GP} tells us that, for large $u$,   $\Pr(Y_1 >u) \approx \alpha u^{-1/\xi}$ for some constant $\alpha$. 
Similarly,  $\Pr(Y_1+...+Y_d >u) \approx c_d u^{-1/\xi_d}$.
If the shape parameter $\xi_d$ decreases with $d$, i.e. $\xi_d < \xi$, then one can always find a large $u$ such that $\alpha u^{-1/\xi} > c_d u^{-1/\xi_d}$. 
This implies that  inequality \eqref{eq: constraint} does not hold for large $u$, and  unwelcome crossing of return levels will eventually occur for large extremes. }

Interestingly, probability theory  confirms this.  For large $u$,  
the so-called Feller approximation \cite{feller:1951} 
states
\begin{equation}\label{eq:Feller}
    \Pr(Y_1+...+Y_d >u) \approx \Pr(Y_1 >u) + \dots + \Pr(Y_d >u) 
\end{equation}
when $Y_i$ are independently and identically distributed (i.i.d.) with a heavy tail behavior. 
Note that extensions with respect to the simple setting of i.i.d. random variables exist in the literature \cite[see, e.g.][]{fougeres:mercadier:2012}.
Practically, this probability result means that 
aggregated precipitation extremes cannot, at least statistically,  become lighter with increasing scales.

In this context, one can wonder why such a strong disagreement between  the practical consensus and the theoretical  aspect appears. 
One main objective of this work is to address this issue and 
to propose a statistical model that can reconcile the theory with practical findings.

Intuitively, from a probability point of view, there are two opposite forces at play when summing. The central limit theorem ``flattens'' the upper tail behavior, but the Feller approximation imposes a constant value of the shape parameter, $\xi$, to avoid high return level crossings.
Additionally, aggregation reduces sample sizes. Consequently, extremal exceedances at scale $d$ are fewer than those of the original data. This makes the generalized Pareto distribution (GPD) approximation less reliable.

Hence, our strategy to avoid this problem is to bypass the threshold selection step.
A fundamental aspect of our approach is to strongly control low and large tail behaviors at all aggregation scales, while keeping the bulk and the transition to tails very flexible.  
This will allow to model a wide range of shapes for the bulk;  from Pareto type like  in the top panels of Figure \ref{fig:expdf} to more Gaussian shapes like in the bottom right panel of Figure \ref{fig:expdf}, while preserving constant tail  shape parameters. 
To do so, we choose a specific distribution class for $Y_i$ and study how the mathematical properties of this distribution change with aggregation.

In this work, we focus on the extended generalized Pareto distributions (EGPD) class \cite{Naveau2016,PAPASTATHOPOULOS2013131}. 
In addition to being a flexible class of distributions, it has been shown to be applicable to various hydrological setups.
For example, the entire distribution of rainfall amounts was modeled using an EGPD in \cite{Gamet22,evin2018}.  
The EGPD class had been   integrated into a random forest scheme in order to improve the post-processing of forecasted rainfall \cite{taillardat2019,taillardat2020}.
It  has also been used to perform  rainfall  comparison
\cite{rivoire21,Rivoire22} and to produce  regional clustering analysis 
\cite{Legall2022}. Besides rainfall analysis with heavy tails, it is also possible to tailor  the EGPD  class  to  model lighter and even bounded variables, such that wind data \cite{turkman2021},  wave heights \cite{Legrand23} or temperatures \cite{Gamet22}.
From an inferential point of view,   EGPD regression approaches have been also studied in detail, \cite[see, e.g.][]{Carvalho22} for a Bayesian hierarchical scheme 
or \cite{Carrer2022} within a distributional regression framework. 
Still, we were not able to find articles that study  the capability of the EGPD  to model theoretically sum of EGPD distributed random variables, and to apply such distributions within a rainfall aggregation context. 

Concerning IDF curves,  various definitions exist \cite[see, e.g.][]{koutsoyiannis1998mathematical}. In most setups, they share two  common threads:  the computation of total rainfall accumulation sums (or averages) over a given time scale $d$ \cite[see, e.g.][]{haruna2023modeling} and  sometimes taking a block maximum, typically over a year or a season \cite[see, e.g.][]{Ulrich20}. 
Mathematically, such operations can be viewed as the transformation of  the original rainfall measurement times series $\bY=(Y_1,Y_2, \dots)^\top$ into a single index. 
For example, a common form of such indices \cite[see, e.g.][]{Ulrich20} is  
\begin{equation}\label{eq:Td}
    T(\bY)= \max\left( \overline{Y}_{1}, \overline{Y}_{2}, \dots \right)
\end{equation}
where $\overline{Y}_{i}$ represents any type of averages over $d$ time steps and the maximum block size corresponds to    the number of such averages over the full period of interest, classically one year. 

In this work, our first  task is to provide a general result, see Proposition \ref{prop: T(X) EGPD},  that details the conditions under which, given that the original time series $\bf Y$ have   identical EGPD marginals,  the aggregated vector   $T(\bY)$ stays EGPD distributed.  
A key element of this result is to understand  how  the constraint \eqref{eq:Feller} can be generalized to $T(\bY)$.
In addition, it will be important to determine how  EGPD parameters change with the transformation $T(\bY)$, to explain how to   infer them.
A limiting aspect of some existing IDF  approaches is that they cannot produce 
curves for short return periods. This is particularly true for IDF methods based on block maxima. 
For example, taking  annual maxima at various aggregation scales, like in \cite{Ulrich20}, 
prevents the computation of any return period below the annual scale. 
To solve this issue, we avoid taking block maxima in our application and, in this paper,  we will mainly  focus  
on the following two  additive aggregation types. Firstly the classical aggregation scheme  \cite[see, e.g.][]{haruna2023modeling}  is defined by 
\begin{equation}\label{eq: Td =sum}
    T(\bY)= Y_1+\dots + Y_d. 
\end{equation}
The second corresponds to a stochastic representation based on the number of wet events over the duration $d$, say $N$ and their corresponding intensities $Y_i$, in the following way
\begin{equation}\label{eq: Td= Pois comp}
    T(\bY)= \sum_{i=1}^{N}Y_i,
\end{equation}
with the convention $T(\bY)= 0$ if $N=0$. 
This stochastic representation is well-known in hydrology \cite[see, e.g.][]{revfeim1984initial,thompson1984homogeneity,dunn2004occurrence,hasan2011two,yunus2017modelling,dzupire2018poisson} and will be a key element in the statistical model proposed in Section
\ref{sec: fitted model}.

Concerning multiscale rainfall modeling, stochastic rainfall generators offer a different avenue. For example, randomized Bartlett-Lewis pulse models \cite[see, e.g.][]{PARK2021} aimed  at reproducing and combining different storm variables (arrival time, duration, intensities, etc).  
Although efficient at reproducing  mean rainfall statistics, underestimation of extreme rainfall may appear. For this reason, we do not pursue this approach here as one of our main focus is to capture extremes distributions at all aggregation scales with statistical guaranties.  Complementary scaling and multifractal approaches provide valuable descriptive insights into cross-scale behavior, but are not usually formulated as generative, likelihood-based models for the full distribution of aggregated rainfall intensities at each duration \citep{Koutsoyiannis2003}

The organization of this article is as follows:  
Section \ref{sec:egpd} recalls the main features of the EGPD  and contains Proposition \ref{prop: T(X) EGPD} which explains how the EGPD parameters can change with aggregation scales. 
In addition, important links between  EGPD and  Poisson compound distributions are investigated.  
This leads to the definition of our statistical model for aggregated rainfall in Section \ref{sec: fitted model}.
 Inference and application are presented in Section 
\ref{sec: inference and application}. Conclusions are given in Section~\ref{sec:conclu}. 
 Proofs of all  propositions, details of a numerical algorithm for evaluating distribution functions, and results of a simulation experiment can be found in the Appendices.

\section{Extended generalized Pareto distributions}\label{sec:egpd}

The fundamental feature of ${H}_{\xi}(\cdot)$ defined in \eqref{eq: GP} 
is its stability under thresholding (up to a normalizing constant). 
Although mathematically sound for modeling heavy rainfall, there is no reason that such a GPD will fit well low and moderate precipitation. 
This limitation has been addressed, and there exists today a wide range of possible statistical options to model the probability distribution of the entire spectrum of precipitation \cite[see, e.g.][]{MacDonald11,Boutigny23}.

\begin{definition}\label{def: MGPD}
Let $U$ be a uniformly distributed random variable on $[0,1]$. Let $B(\cdot)$ be a cumulative distribution function (cdf) on $[0,1]$ with a continuous probability density function (pdf) $b(\cdot)$ on $(0,1]$ and a possible non-zero mass $B(0)$ at $0$.
The non-negative random variable defined as 
\begin{equation}\label{eq: Y and B}
  Y = \sigma H_{\xi}^{-1}\left( \left(B^{-1}\left( U\right)\right)^{1/\kappa} \right),
\end{equation}
where $\sigma, \kappa, \xi$ are three positive constants,
is said to follow an extended generalized Pareto distribution (EGPD), denoted by $Y \sim EGPD(\sigma,\kappa, \xi, B)$, if 
\begin{equation}\label{eq: constraints on b}
  0< b(0^+) < \infty \mbox{ and } 0< b(1) < \infty,
\end{equation}
where  
\begin{equation}\label{eq: conditions zero}
 b(0^+) = \lim_{u \rightarrow 0^+} \frac{B(u)-B(0)}{u}
\end{equation}
\end{definition}

The definition \eqref{def: MGPD} is based on that provided by \cite{Naveau2016}, but differs in terms of notation, 
as the cdf $G(u)=B(u^\kappa)$ was used in \cite{Naveau2016}.
The notation with $G(\cdot)$ was slightly ambiguous because the parameter $\kappa$ driving the lower tail was "hidden" in the cdf G itself, which was not the case for $\xi$. In contrast, the new definition, $EGPD(\sigma, \kappa, \xi, B)$, more precisely distinguishes the roles of $\kappa$ in modeling the lower tail, $B(\cdot)$ as the transfer function from low to heavy intensities, and $\xi$ in modeling the upper tail.
Although this work does not explicitly model dry events, we recognize their important role in aggregated rainfall distributions. 
In this respect, Definition ~\ref{def: MGPD} is further different from the definition in \cite{Naveau2016}. It separates the probability of a dry event, $B(0)$, and the behavior of the distribution of $Y$ for small but positive values near $0$.

More formally, one can check that the lower tail satisfies 
\begin{equation}
\label{eq:lower}
P(Y\leq y) =B(0) + b(0^+) \left(\frac{y}{\sigma}\right)^{\kappa} + o(y^\kappa) \quad \mbox{as $y \rightarrow 0^+$},
\end{equation}
and the upper tail is GP equivalent, i.e.
\begin{equation}
\label{eq:upper}
P(Y> y) =\kappa b(1) \overline{H}_{\xi}(\frac{y}{\sigma}) + o(y^{-\frac 1 \xi})  \quad \mbox{as $y \rightarrow + \infty$}
\end{equation}
with $\overline{H}_{\xi}(\frac{y}{\sigma}) = 1-H_{\xi}(\frac{y}{\sigma})$. 
The pdf and cdf expressions the EGPD, as well as the proof of \eqref{eq:lower} and \eqref{eq:upper}, can be found in \ref{sec:EGPD}.

The special case $B(u)=u$ is called Type 1, see the case $G(u)=u^\kappa$ with $\kappa \neq 0$ in \cite{Naveau2016}. 
It offers a flexible choice to model the full range of hourly and daily rainfall, extremes included, see \cite{blanchet2018,evin2018,li2012}. When $\kappa=1$, it corresponds to the classical generalized Pareto distribution. Another member of the EGDP class used in hydrology is the so-called Pareto-Burr-Feller proposed in \cite{Koutsoyiannis2017}.

\cite{haruna2023modeling} used a EGPD of Type 1 to derive an IDF for different aggregation scales.
To add flexibility, they allowed the shape parameter $\xi$ to vary with the aggregation scale. As mentioned earlier, this leads to a mathematical contradiction with \eqref{eq:Feller} and allows for unwanted return level crossings in extremes.
One option to be in compliance with \eqref{eq:Feller} is to fix the shape parameter $\xi$ but then, to fit the data adequately, the function $B(u)$ needs to vary with the aggregation scale. 
We will follow this modeling path that leads to the question of how to characterize mathematically 
$B(u)$ changes with the aggregation scale. Before answering this complex question, we need to derive some properties of the EGPD.

\subsection{Sums and other transformations of EGPD random variables}
A fundamental question regarding IDF curves is to determine how the distributional features of a stationary time series, say $\bY=(Y_1, Y_2, \dots)^\top$, change according to a transformation $T(\cdot)$. When all $Y_i$ have the same EGPD marginal, one can wonder what are the sufficient conditions to ensure that the real-valued transformed vector $T(\bY)$ has also a EGPD distribution, not necessarily with the same parameters as projecting the data at hand into a single value index will likely change the bulk of the distribution. For example, the sum of two type 1 EGPD is not a type 1 EGPD, but we will show that it is still a EGPD. 
The following proposition provides the precise conditions to ensure this. 
Note that this result is broad enough to allow  dependencies among the $Y_i'$s, and to avoid imposing a specific  form on the transform $T(\cdot)$.

\begin{proposition}\label{prop: T(X) EGPD}
Let   $T$ be a non-negative random variable and $Y \sim EGPD(\sigma,\kappa, \xi, B)$  for   $\xi>0$, $\kappa>0$.
If  there exist some positive and finite constants $\alpha$, $\beta$ and $\gamma$ such that   
\begin{equation}\label{eq: tail equivalence conditions upper T}
    \lim_{y \to \infty } \frac{\Pr( T > y) }{\Pr( Y> y) }=\alpha,
\end{equation}
\begin{equation}\label{eq: tail equivalence conditions lower}
            \lim_{y \to  0^+ } \frac{\Pr( 0<  T  \leq  y) }{[\Pr(0< Y \leq  y) ]^\gamma}=\beta, 
\end{equation} 
then 
 there exists a  cdf $B_T$ such that
$T \sim EGPD(\sigma,\gamma \kappa, \xi, B_T)$
with 
\begin{equation}\label{eq:b_T}
b_T(0^+)=  \beta \; b(0^+)^\gamma  \mbox{ and } b_T(1)= \alpha \; b(1)/\gamma.
\end{equation}
\end{proposition}

Condition \eqref{eq: tail equivalence conditions upper T} holds for most applications when $T=T(\bY)$ with $T(\cdot)$ one of the usual transforms like 
\eqref{eq:Td}, \eqref{eq: Td =sum} or \eqref{eq: Td= Pois comp} applied when working with aggregated data.
It forces the upper tail of $T(\bY)$ to be proportional to the marginal one.
In cases like \eqref{eq: Td =sum}, it corresponds to Feller's lemma, see \eqref{eq:Feller}. 
For transforms like 
\eqref{eq: Td= Pois comp}, it is related  to Breiman's lemma \cite{breiman:1965}.  It basically tells us that the upper tail behavior of a  random sum of heavy tailed distributed random variables is driven by the largest tail index. \cite{fougeres:mercadier:2012} detailed different setups for dependent cases and \cite{ jessen:mikosch} reviews existing results for a variety of transformations $T(.)$, including the case
   $ T(\bY)= \max (Y_1,\dots,Y_d)$.
Condition \eqref{eq: tail equivalence conditions lower} does the same but for the lower tail. 
Checking the validity 
 of condition  \eqref{eq: tail equivalence conditions lower} 
is also possible for simple setups.
For example, we detail the i.i.d. case in \ref{sec:appendixlow} with five different setups. The case $\gamma>1$ appears when $B(0)=0$ (only wet events), while $\gamma=1$ otherwise. In the remainder of the paper, this later case is assumed, meaning that the behavior of the lower tail of $T(\bY)$ is proportional to that of  $Y_i$.

An interesting practical outcome of this proposition is that the shape parameters, $\kappa$ and $\xi$, which are related to the lower and upper tails of the distribution, remain unchanged when the transform $T(\cdot)$ is applied.  
Equation \eqref{eq:b_T} also provides the values of $b_T(u)$ for $u = 0^+$ and $u = 1$. 
It would be of  great interest to also   deduce, for all $u$ in $(0, 1)$,  the function $b_T(u)$ in function of $B(u)$ for any $d$ for the simple aggregate defined by \eqref{eq: Td =sum}.
If possible, then the practitioner could just focus on modeling the finest time scale to infer $B(\cdot)$. 
But, this task is mathematically challenging. 
For example, suppose that at the finest timescale (six minutes in our application), the non-zero observations simply follow an i.i.d. GPD sample. 
Even in this simple case, we were unable to determine the parametric form of $b_T(u)$ for any $u\in(0,1)$. Consequently, the pdf of $T(\bY)=Y_1+\dots+Y_d$ is not explicit.

In this work, compound sums \eqref{eq: Td= Pois comp} are used as a model for aggregated sum \eqref{eq: Td =sum}. This modeling choice is motivated by interpretability, theoretical and computational reasons, which are detailed in the next section.

\subsection{Compound Poisson-EGPD distributions}
\label{se: CEGPD}

A   classical modeling approach in hydrology considers that rainfall time series can be decomposed into a succession of rainfall events, which arise as a point process with rainfall amounts associated to each of these events \cite{rodriguez1987}. 
A natural  stochastic representation of aggregated rainfall associated to such representation was recalled in the introduction via the compound sum  \eqref{eq: Td= Pois comp} \cite[see, e.g.][]{revfeim1984initial,thompson1984homogeneity,dunn2004occurrence,hasan2011two,yunus2017modelling,dzupire2018poisson}. 
To make the link with the EGPD class, we introduce  the following definition.

\begin{definition}
\label{def:CEGPD}
Let $\bY^*=(Y^*_1,Y^*_2, \dots)^\top$ 
be a sequence of  i.i.d.~EGPD random variables,  $Y^*_i \sim EGPD( \sigma,\kappa,\xi,B)$, with $B(0)=0$,  $\kappa>0$ and $\xi>0$.  Let $N$ be  a Poisson random variable  with mean $\lambda$ independent of $\bY^*$. 
The distribution of the  random sum defined by  
$    
    \sum_{i=1}^{N} Y^*_i,
$
with the  convention that $\sum_{i=1}^{0} Y^*_i=0$, is called  a compound Poisson-EGPD. 
\end{definition}

The following proposition shows that Compound Poisson-EGPD is a sub-family of EGPD. Combined with Proposition ~\ref{prop: T(X) EGPD}, it implies that compound Poisson-EGPD have the same tails that aggregated sums \eqref{eq: Td =sum} of EGPD random variables. For mathematical completeness, Proposition~\ref{lem: CEGPDv2} characterizes more precisely Compound Poisson-EGPD as the sub-family of  EGPD which are infinitely divisible, i.e. that can be expressed as the probability distribution of aggregated sums \eqref{eq: Td =sum} of any arbitrary number $d$ of i.i.d. random variables.

\begin{proposition}\label{lem: CEGPDv2}

A compound-EGPD random variable is an EGPD random variable. More precisely, with the notations of Definition~\ref{def:CEGPD}, we have

$$\sum_{i=1}^{N} Y^*_i \sim  EGPD( \sigma,\kappa,\xi, B_{\lambda})$$ with $
B_{\lambda}(0)=\exp(-\lambda) $, $ b_{\lambda}(0^+)=\lambda  \exp(-\lambda)  b(0^+)$ and  $b_{\lambda}(1)=\lambda b(1)$.

Conversely, any infinitely divisible $EGPD(\sigma,\kappa,\xi,B)$ random variable with $B(0)>0$ is a compound Poisson-EGPD random variable.
\end{proposition}

At this stage, let us highlight that the random variables $Y_i^*$ in  Definition \ref{def:CEGPD}  can be very general as the cdf $B$ has only the requirements \eqref{eq: constraints on b} and \eqref{eq: conditions zero}. To balance between parameters parsimony and model versatility,  we fix $B(u)=u$ and let $\lambda$ vary according to the aggregation scale. This leads to the following definition.

\begin{definition}
\label{def:CEGPD+}
In the particular case when $B(u)=u$, the positive\footnotemark{} part of the distribution of the random sum $\sum_{i=1}^{N} Y^*_i$ in Definition~\ref{def:CEGPD} is denoted $EGPD( \sigma,\kappa,\xi, \lambda)$.
\end{definition}

Proposition~\ref{lem: CEGPDv2} implies that if
$Y \sim EGPD( \sigma,\kappa,\xi, \lambda)$ then there exists a cdf $B_\lambda$ such that $Y\sim EGPD( \sigma,\kappa,\xi, B_\lambda)$ with explicit expressions for $b_\lambda(0^+)$ and $b_\lambda(1)$. Remark that $B_{\lambda}(0)=0$ since only the positive part of the compound Poisson-EGPD is kept in Definition~\ref{def:CEGPD+}. However the bulk of the cdf $B_{\lambda}$ and the pdf of $Y$ are unknown. This issue is addressed by observing that the pdf of compound sums of the form \eqref{eq: Td= Pois comp} can be numerically approximated with a low computational cost (see e.g. \cite{embrechts2009panjer} and references therein). The numerical results given in this paper were obtained using the Panjer's algorithm \cite{panjer1981recursive} detailed in \ref{sec:Panjer}.

\begin{figure}[t]
    \centering
    \includegraphics[scale=.75]{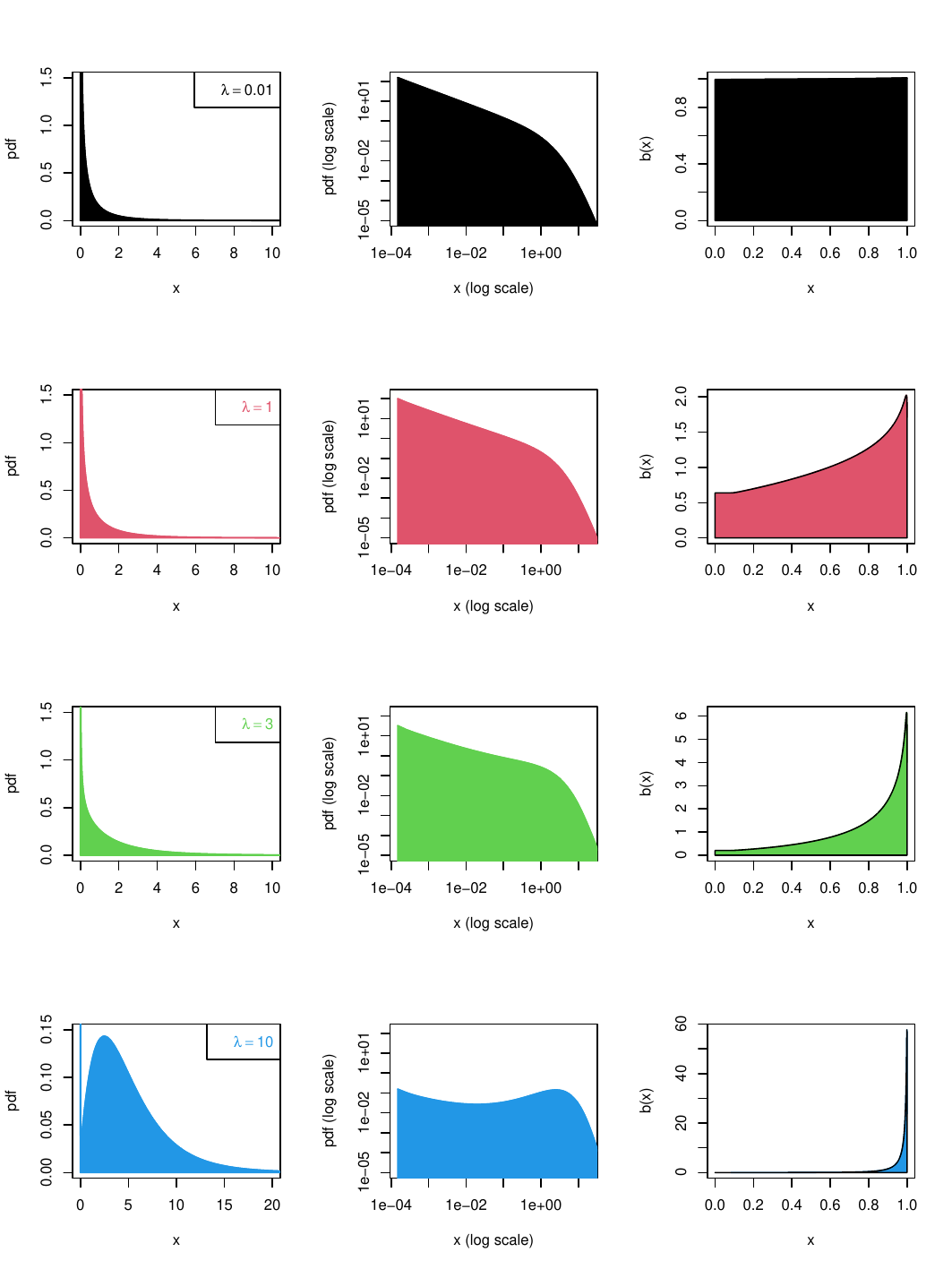}
    \caption{Left panels: pdf  of the $EGPD(\sigma,\kappa,\xi,\lambda)$, see Definition \ref{def:CEGPD+},  with $\kappa=0.3$,  $\sigma=1$, $\xi =0.25$  and various values of the mean $\lambda$ given in the legend. Middle panels: same as left panels with log scale on both axis.
    Right panels: corresponding pdfs $b_\lambda(\cdot)$.  All the pdfs are numerically approximated  using the Panjer recursions, see \ref{sec:Panjer}}
    \label{fig:exCEGPD}
\end{figure}

When all parameters except for the Poisson mean $\lambda$ are fixed,  $\lambda$ is the sole factor that drives the transition from the lower to the upper tail of the $EGPD( \sigma,\kappa,\xi, B_\lambda)$.
 The right panels of Figure~\ref{fig:exCEGPD}  shows the function $b_\lambda(\cdot)$ for $\lambda \in \{0.01, 1,3,10\}$   (from top to bottom panels). In these plots, the other parameters are equal to $\kappa=0.3$, $\sigma=1$ and $\xi=0.25$. These are typical values for the rainfall data considered in this study (the shapes of the pdf plotted on Figures~\ref{fig:expdf} and \ref{fig:exCEGPD}  share similarities). The corresponding pdfs are displayed in the left panels. To emphasize the log-linear behaviors in the tails, the middle panels show these pdfs in log scale for the x- and y-axis, these slopes being related to the parameters $\kappa$ and $\xi$, see  Proposition~\ref{lem: CEGPDv2}.

In the top row, when $\lambda$ is close to zero, we can recognize a type 1 EGPD  as $b(u) \approx 1$.
This can be retrieved formally by computing the limit of $\Pr(N=n|N>0)$ when $\lambda \rightarrow 0$ for a Poisson random variable $N$, which is equal to $1$ for $n=1$ and $0$ for $n \geq 2$. 
For the largest values of the parameter $\lambda$ (see the last row in blue), the pdf of the EGPD is bimodal, with a sharp mode at $0$ and a second mode in the bulk of the distribution, i.e. the central limit theorem  appears here.

\section{A statistical model for aggregated rainfall}\label{sec: fitted model}

Let $D$ be the largest scale of aggregation under study. For any integer $d=1,\dots,D$, we 
denote 
\begin{equation}\label{eq:amount}
A_d=(Y_1+\dots+Y_d)^+
\end{equation}
the random variable which describes the positive\footnotemark[\value{footnote}]
\footnotetext{
Both the compound-EGPD and the rainfall distributions have a point mass at zero. However, numerical experiments have shown that additional challenges arise when attempting to simultaneously model dry and wet conditions. For example, an excess of zeros in the rainfall distribution necessitates zero-inflated distributions. Additionally, different temporal dynamics in dry and wet events affect the aggregation properties. For this study, only the positive part of the rainfall distributions is modeled.}  precipitation amount at aggregation scale $d$, where zero values are truncated after computing aggregated rainfall $Y_1+\dots+Y_d$.

In the preceding section, it was pointed that the compound EGPD can serve as an appropriate statistical model for the distribution of aggregated rainfall data.
 In the remaining of the work, we thus assume  that for any $d=1,\dots,D$
\begin{equation}\label{eq:mod}
    A_d \sim  EGPD(\sigma_d,\kappa,\xi,\lambda_d).   
\end{equation}

	It should be noted that the scale parameter, $\sigma_d$, and the mean,  $\lambda_d$, are allowed to vary with $d$, while the parameters  $\kappa$ and $\xi$  are assumed to be constants.
	In this study, we hypothesize that the following parametrization is applicable:
	\begin{equation}
		\label{eq:modpara}	
		\log \sigma_d = \sum_{i=0}^p s_i \left(\log d \right)^i \ \ , \ \    \log\lambda_d=\sum_{j=0}^q l_j \left(\log d \right)^j.
	\end{equation}
Such parametrization share similarities with   classical models used in the literature on IDF curves. For example, log-linearly changes in the    scale  parameter were present in the 
GEV approach based on  annual block  maxima  \cite[see, e.g.][]{ulrich2021modeling}, or the GPD approach when dealing with    threshold exceedances \cite[see, e.g.][]{van2010construction}. 
A significant difference between our EGPD modeling and the one in \cite{haruna2023modeling} is that our shape parameters $\kappa$ and $\xi$ remain constant across different aggregation scales, to be consistent with Propositions~\ref{prop: T(X) EGPD} and \ref{lem: CEGPDv2}.

The coefficients $s_i$ and $l_j$ are not variation free. Proposition~\ref{prop:monotony} implies that the non-crossing return level condition \eqref{eq: constraint}, which is an important motivation for this work, is true when both functions $\sigma_d$ and $\lambda_d$ increase with $d$. These monotonicity conditions will be used as constraints in the estimation procedure.

\begin{proposition}
    \label{prop:monotony}
   Let $Y_1,Y_2,\dots$ be a stationary sequence of non-negative random variables
such that assumption \eqref{eq:mod} is satisfied with $\sigma_d \geq \sigma_1$ and $\lambda_d \geq \lambda_1$. Then \eqref{eq: constraint} holds true.
\end{proposition}

\section{Inference and application}\label{sec: inference and application}

\subsection{Parameter estimation}

The model introduced in Section~\ref{sec: fitted model} has $p+q+4$ parameters encapsulated in the vector  $\theta=(s_1, \dots, s_p,\kappa,\xi, l_1,\dots,l_q)$.  
Although precipitation data are generally not independent in time and across aggregation scales,  we estimate these parameters 
 by maximizing the   following   composite likelihood function $L(\theta)$,
constructed under working independence assumptions \cite[see, e.g.][]{chandler2007inference},
\begin{equation} 
\label{eq:CL}
L(\theta)=\prod_{d =1}^D \prod_{i=1}^{n_d} p(a_{d,i};\sigma_d,\kappa,\xi,\lambda_d)
\end{equation}
where $(a_{d,1},\dots,a_{d,n_d})$ denotes the sample of positive rainfall amounts available at aggregation scale $d$ (see \eqref{eq:amount}), $p(\cdot;\sigma,\kappa,\xi,\lambda)$ the pdf of the $EGPD( \sigma_d,\kappa,\xi, \lambda_d)$ (see \ref{sec:Panjer} for more details) and $\sigma_d$ and $\lambda_d$ are function of $\theta$, see \eqref{eq:modpara}.

The computational cost of maximizing  \eqref{eq:CL} can be significant if we consider a large number of aggregation scales, $D$. 
In such cases, an estimate of $\theta$ can be obtained using a smaller subset of $\{1, \ldots, D\}$ without a significant loss of efficiency.
In this paper, we maximize \eqref{eq:CL} with the subset  $d\in \{1,2,...,10,20,...,240,270,...,720\}$, i.e. 
instead of focusing on 720 durations, we focus on 42 representative durations. These durations correspond to all available sub-hourly durations and multiples of hourly durations up to the daily scale and on multiples of three hourly durations up to three days. 

Regarding  confidence bands around our estimates from \eqref{eq:CL}, we follow   \cite{haruna2023modeling}
by using block bootstrap. All the numerical results presented in this study were  obtained using a block size of two  weeks length  and by repeating the optimization procedure on $500$ bootstrap samples.
We validated the estimation procedure using simulations in an idealized i.i.d. setting. The results can be found in \ref{sec:Appendixsimu}.

\subsection{Rainfall analysis of French precipitation intensities}
\label{sec:data}

\begin{figure}
    \centering
    \includegraphics[width=13cm]{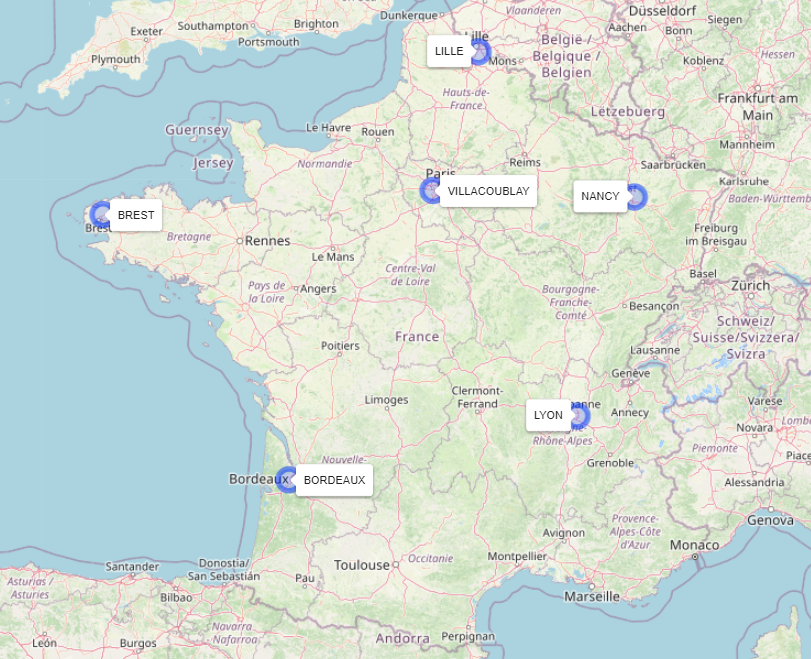}
    \caption{Locations of the meteorological stations considered in this study. 
    M\'et\'eo-France recorded  6-minute rainfall time series for  the period 2006-2023 (with a tipping bucket precision of $0.2$ mm).}
    \label{fig:carte}
\end{figure}

In this study, we examine rainfall data recorded by M\'et\'eo-France at six meteorological stations in France, see  Figure ~\ref{fig:carte}, that represent different climate.
Data are available at the url \url{https://meteo.data.gouv.fr/datasets/donnees-climatologiques-de-base-6-minutes/}  
with a 6 min time step from 2006 until 2023. 
To remove the seasonal component, the focus is on August and September, when intense convective precipitation events typically occur. Summary statistics of 6-minute rainfall data by station are given in Table
\ref{tab:stations}.
All data were obtained using tipping bucket gauges with 0.2 mm precision.  
To compute the composite likelihood \eqref{eq:CL}, the same 0.2~mm discretization is applied to the distribution of the seed before running the Panjer recursions, see \ref{sec:Panjer}.

\begin{table}[ht]
\centering
\begin{tabular}{lccccc}
\hline
\makecell{Station} &
\makecell{Altitude\\(m)} &
\makecell{Percentage\\of missing\\values} &
\makecell{Percentage\\of positive \\data} &
\makecell{Mean \\ of positive \\data (mm)} &
\makecell{St. dev. \\of positive \\data (mm)} \\
\hline
BREST        &  92 & 0.12 & 3.04 & 0.33 & 0.37 \\
NANCY        & 212 & 0.00 & 2.34 & 0.38 & 0.52 \\
LILLE        &  47 & 0.50 & 2.25 & 0.38 & 0.61 \\
BORDEAUX     &  47 & 0.03 & 1.59 & 0.44 & 0.71 \\
VILLACOUBLAY & 174 & 1.28 & 1.71 & 0.38 & 0.52 \\
LYON         & 235 & 0.01 & 2.09 & 0.44 & 0.71 \\
\hline
\end{tabular}
\caption{Summary statistics of 6-minutes rainfall data by station. Results for August-September based on 18 years of data.}
\label{tab:stations}
\end{table}

\begin{figure}[t]
    \centering
    \includegraphics[height=6cm]{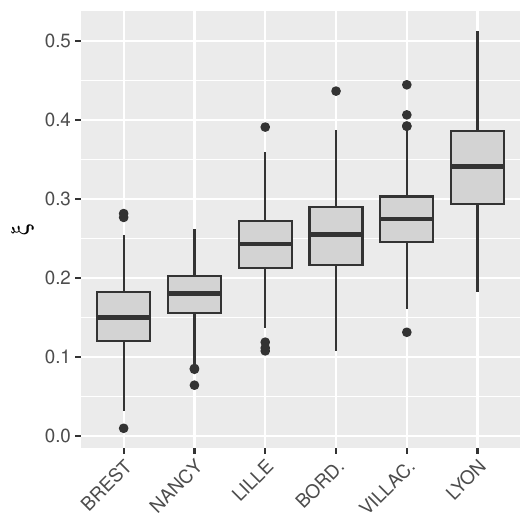}
    \includegraphics[height=6cm]{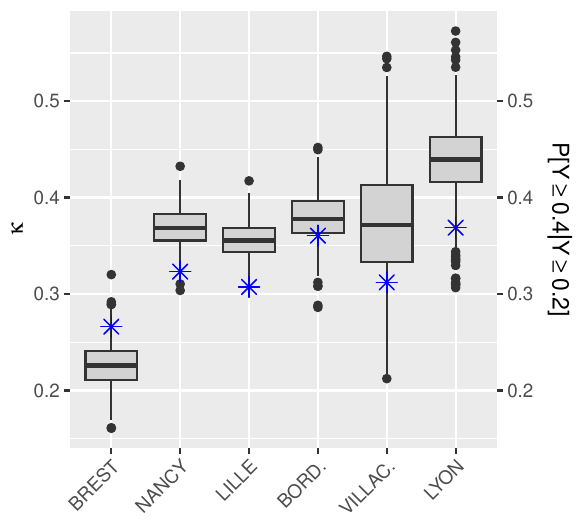}
    \caption{Estimation of $\xi$ (left panel) and $\kappa$ (right panel) of model \eqref{eq:mod} applied to  the eight locations shown in Figure \ref{fig:carte}. The x-axis is ordered with respect to the estimated value of $\xi$. The boxplots are computed using block bootstrap. The blue stars on the right panel represent the empirical estimate of $P[Y\geq 0.4]=1-P[Y=0.2]$ where $Y$ denotes the 6-minutes rainfall data. Results for August-September based on 18 years of six-minutes data.}
    \label{fig:shape}
\end{figure}

The left panel of Figure~\ref{fig:shape} displays the estimate of the upper tail parameter at the different stations considered in this study. The point estimates of $\xi$ seem physically plausible and range from approximately $0.15$ in Brest, close to the Atlantic Ocean, to $0.35$ in Lyon where more intense convective events are observed.  The estimates of the lower tail parameter $\kappa$ are all in the interval $(0.2,0.45)$ (see right panel of Figure~\ref{fig:shape}), corresponding to distributions with a relatively sharp mode at $0^+$. Interestingly, the estimate of $\kappa$ seems to be related to the percentage of measurement greater than $0.2$ in the wet measurements (see the stars superimposed to the boxplot). This is consistent with the interpretation that $\kappa$ describes the lower tail of the distributions and should smaller at stations where light rain conditions are more frequent, as it is the case in Brest for example.

\begin{figure}[t]
    \centering
      \includegraphics[width=6.5cm]{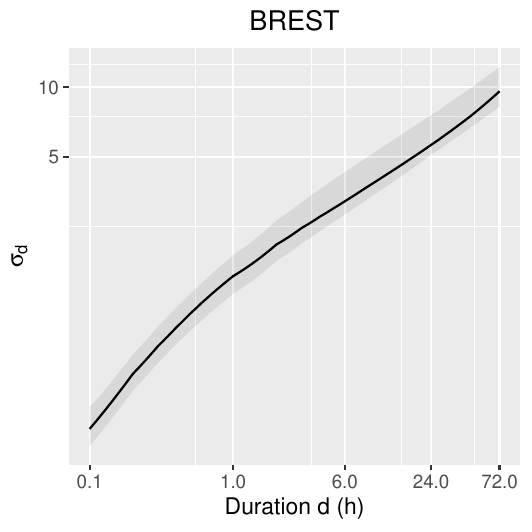}
    \includegraphics[width=6.5cm]{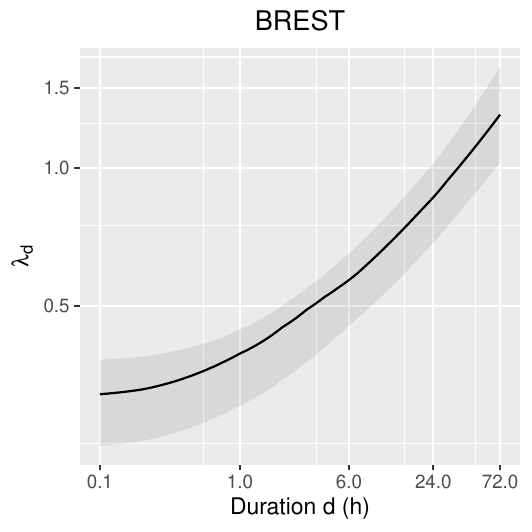}\\
    \caption{Estimation of $\sigma_d$ (left panel) and $\lambda_d$ (right panel) as a function of the duration of aggregation $d$ in Brest. The $95\%$ confidence bands were computed using block bootstrap. Results for August-September based on 18 years of 6-minute data.}
    \label{fig:evolpara}
\end{figure}

Except the differences in the values of $\kappa$ and $\xi$ discussed above, similar fitting results were generally obtained at the different stations. We therefore choose to focus on one of these stations, namely Brest. Figure~\ref{fig:evolpara} shows the evolution of the parameters $\sigma_d$ and $\lambda_d$ with the aggregation scale $d$. As expected, the functions are increasing, since the constraint is imposed in the estimation procedure to ensure the non-crossing condition \eqref{eq: constraint} (see discussion before Proposition~\ref{prop:monotony}). This is also consistent with the physical interpretation of the model, with $\lambda_d$ representing the mean number of rainfall events on the aggregation scale $d$ and $\sigma_d$ describing the scale of their intensities.

\begin{figure}[t]
    \centering
    \includegraphics[width=13cm]{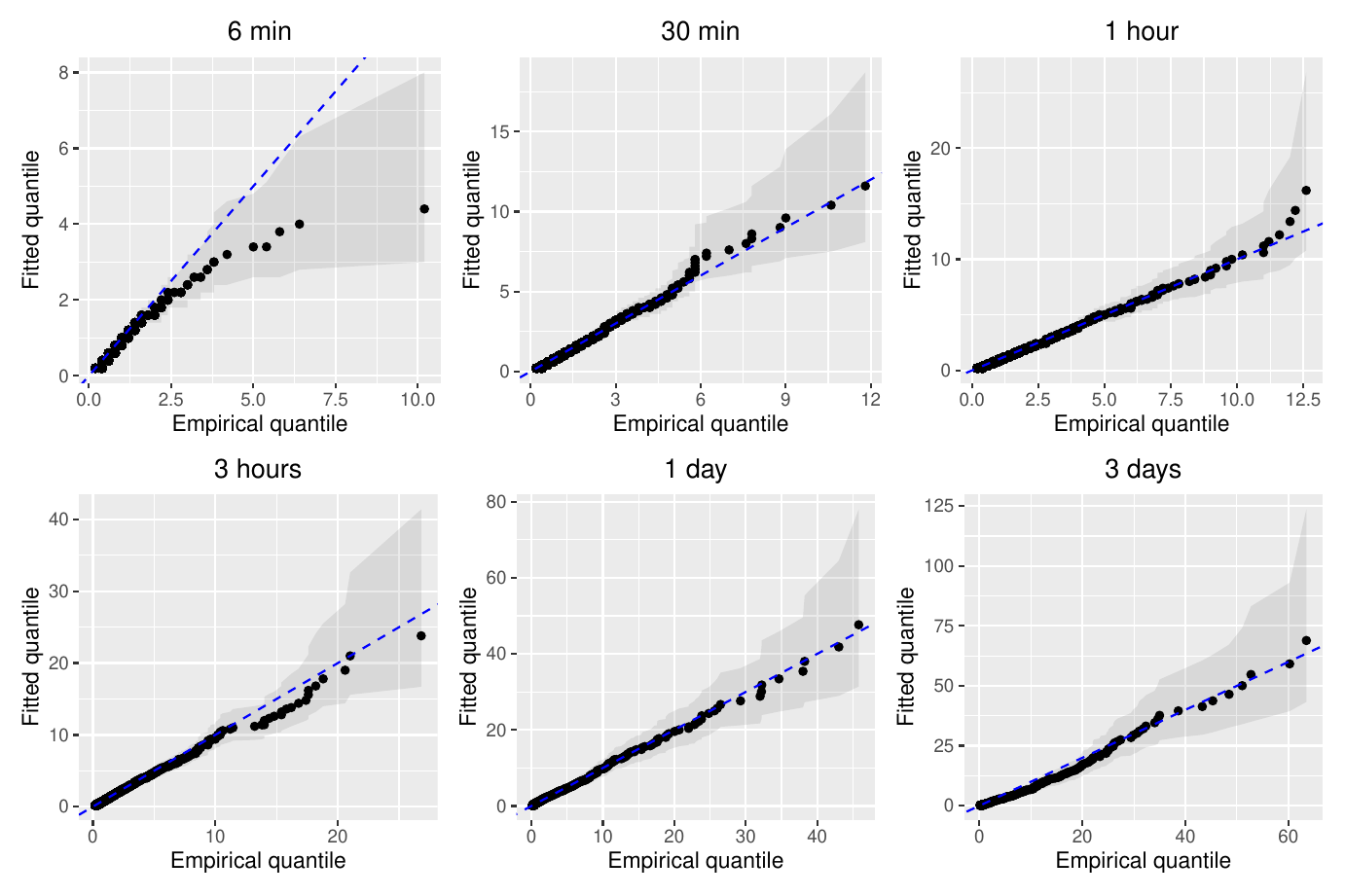}
    \caption{Quantile-quantile plots of the fitted model at different aggregation time in Brest. The $95\%$ confidence bands were computed using block bootstrap. Results for August-September based on 18 years of 6-minute data.}
    \label{fig:qqplotB}
\end{figure}

\begin{figure}[t]
    \centering
    \includegraphics[width=6.5cm]{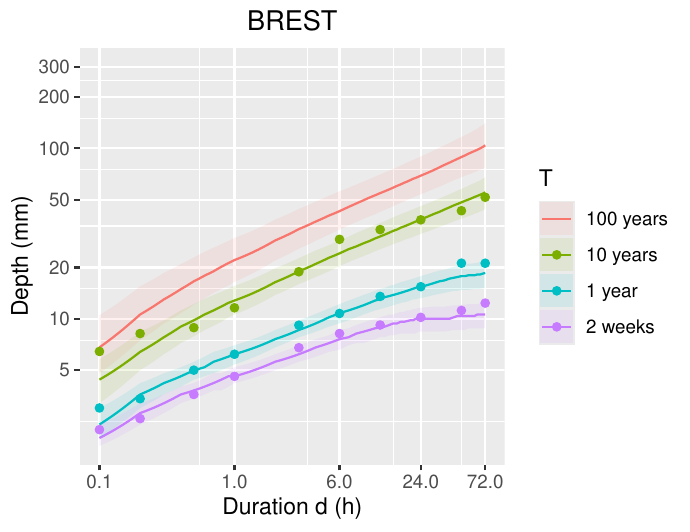}
    \includegraphics[width=6.5cm]{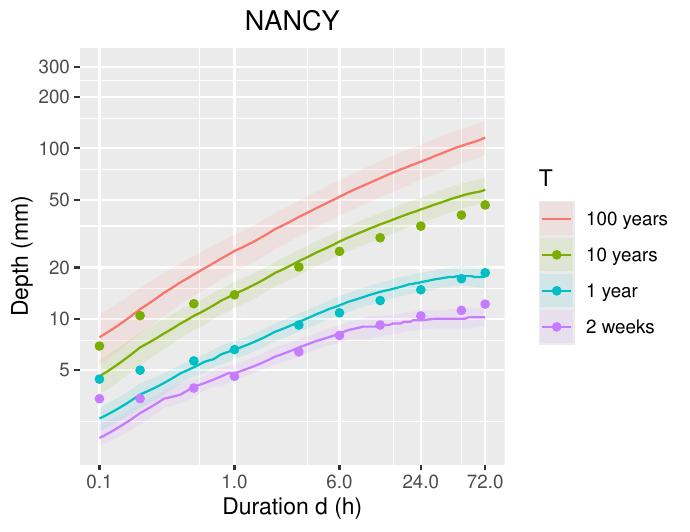}
    \includegraphics[width=6.5cm]{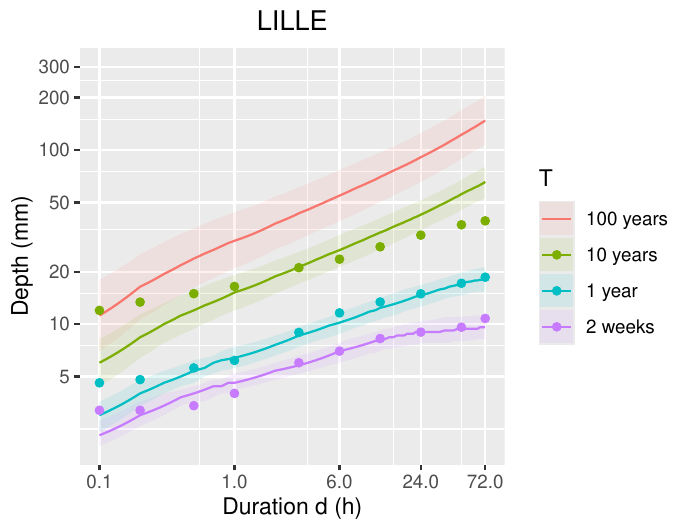}
    \includegraphics[width=6.5cm]{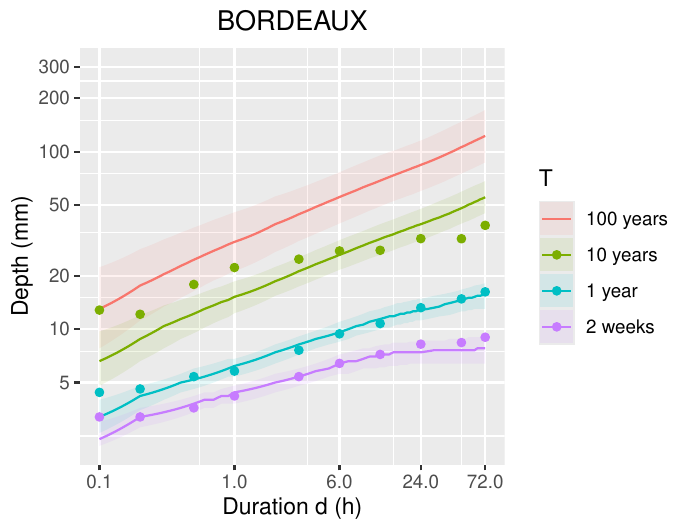}
    \includegraphics[width=6.5cm]{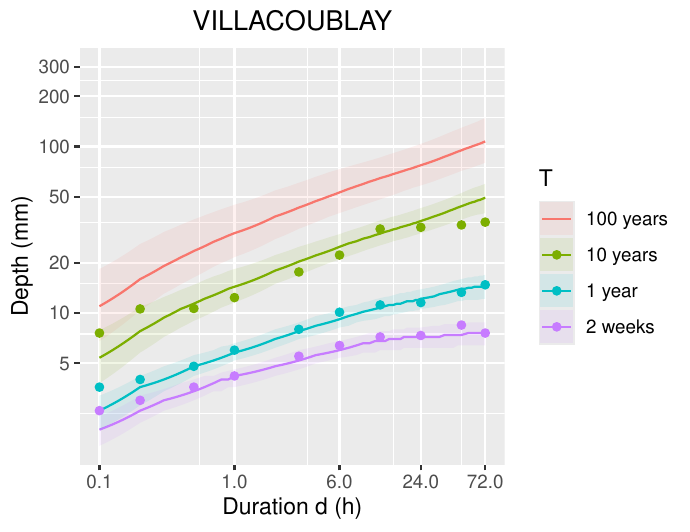}
    \includegraphics[width=6cm]{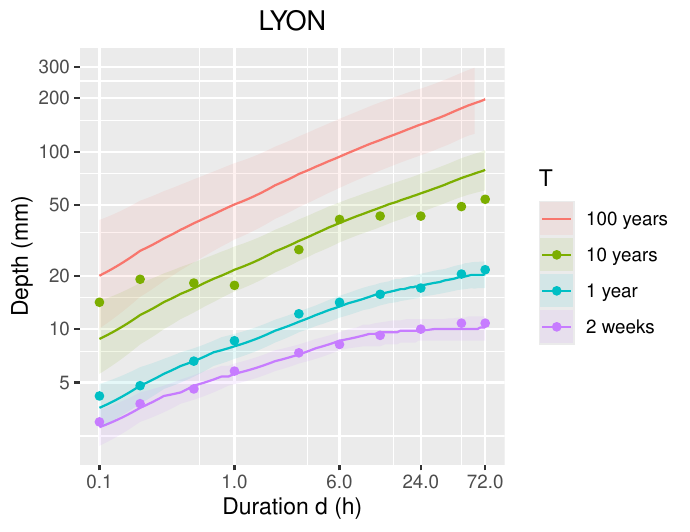}
    \caption{Quantiles of order $1-\frac{1}{n_dT}$ as a function of $d$ with $n_d$ the number of observations available per month at the duration of aggregation $d$. The $95\%$ confidence bands were computed using block bootstrap. The points represent the corresponding empirical quantiles (only for  $T \leq 10$ years). Results for August-September based on 18 years of 6-minute data.}
    \label{fig:quantiles}
\end{figure}

Figure~\ref{fig:qqplotB}  compares the empirical distributions of rainfall with those given by the fitted parametric model at different representative aggregation scales, as illustrated by QQ plots (quantile-quantile plots).
 The model generally fits well for aggregation scales between 30 minutes and three days. However, it slightly underestimates the probability of extreme events at finer scales.
 Figure~\ref{fig:quantiles} provides another representation of the quantiles at different aggregation scales, which illustrates this as well. The figure focuses on high-order quantiles associated with return periods ranging from 15 days to 100 months.
For the six stations examined in this study, the empirical quantiles generally align closely with the theoretical quantiles predicted by the fitted model for aggregation durations ranging from 30 minutes to one day. However, there is a consistent underestimation of larger quantiles for shorter durations and, conversely, a slight overestimation at larger aggregation scales for certain stations.
More flexible models need to be investigated to handle such durations with the proposed methodology. Once again, it is worth noting that, unlike the models usually used to construct IDF curves, the fitted model describes the entire distribution and can therefore be used to compute quantiles associated with short return periods.

\section{Conclusions}
\label{sec:conclu}
A new statistical model has been proposed for the distribution of positive precipitation on various aggregation scales. The model uses the compound Poisson-EGPD as a key ingredient. This distribution was chosen for its physical interpretability and computational advantages. Theoretically, it is also shown to provide a tail-compliant model for the distribution of aggregated data. The proposed model is parsimonious, with only eight parameters, and can be fitted to data at low computational cost using the Panjer algorithm. 
The model was fitted to 6-minute rainfall data from six stations in France 
with varying climates. The entire distribution of positive rainfall data was found to be generally well described by the proposed model at aggregation scales ranging from 30 minutes to three days.

These results are encouraging and outline possible avenues for future research. Firstly, systematic validation across multiple stations and seasons could be considered. Secondly, longer observation periods could be used to take into account possible climate changes. From a model definition point of view, introducing a time-dependent parameterization to the model \eqref{eq:mod} does not present any particular estimation problems. However, it is worth noting that this model is justified on the basis of the results of Propositions \ref{prop: T(X) EGPD} and \ref{lem: CEGPDv2} which are valid for identically distributed variables. Extending these results to non-identically distributed variables is an interesting theoretical problem. Finally, in the case of bounded–tail distributions, an extension of the EGPD framework is possible (see \cite{Legrand23}), but adapting the theoretical results developed in this paper to that setting would require substantial additional work, as the key asymptotic arguments used here no longer apply directly.

\section*{Acknowledgments}
The authors acknowledge the support of the SHARE PEPR Maths-Vives project (France 2030 ANR-24-EXMA-0008).

Part of Gaetan's research work took place within the framework of the DoE 2023-2027 (MUR, AIS.DIP.ECCELLENZA2023\textunderscore27.FF project). 

Part of Naveau’s research work was supported by the  French  Agence Nationale de la Recherche:  EXSTA, the PEPR TRACCS programme under grant number (PC4 EXTENDING, ANR-22-EXTR-0005), and  the PEPR   IRIMONT (France 2030 ANR-22-EXIR-0003). He has also benefited  from the Geolearning research chair, a joint initiative of Mines Paris and the French National Institute for Agricultural Research (INRAE).

\newpage

\appendix

\section{Basic EGPD properties}
\label{sec:EGPD}
The cdf of a $EGPD(\sigma,\kappa, \xi, B)$ can be expressed as 
\begin{equation}\label{eq: F bar}
    F(y)= B\left(H_\xi^\kappa(y/\sigma)\right), \mbox{ for any $y\geq 0$}
\end{equation}
and the pdf of a $EGPD(\sigma,\kappa, \xi, B)$ can be written as 
\begin{equation}\label{eq: pdf}
      \frac{\kappa}{\sigma} \, h_\xi(y/\sigma) 
    \, H^{\kappa-1}_\xi(y/\sigma) \, b\left( H^{\kappa}_\xi(y/\sigma)\right), \mbox{ for $y> 0$,}\\ 
\end{equation}
where $h_\xi(\cdot)$ corresponds to the pdf of a Generalized Pareto random variable.

Since $H_\xi(y)=y +o(y)$ for   $y$ near zero, $h_\xi(0)=1$ and $H_\xi(0)=0$, we have 
\begin{equation}\label{eq: b(0)}
   \lim_{y\rightarrow 0^+}\frac{F(y)-F(0)}{y^\kappa}=\lim_{y\rightarrow 0^+}\frac{f(y)}{\kappa y^{\kappa-1}}= \frac{b(0^+)}{\sigma^\kappa}. 
\end{equation}
where the first relation was derived via L'H\^opital's rule.
This implies \eqref{eq:lower}.

Concerning the upper tail behavior L'H\^{o}spital's rule and \eqref{eq: pdf} give 
\begin{equation}\label{eq:kappa.b(1)}
    \ytoinf \frac{1-F(y)}{1-H_\xi(y/\sigma)}= 
\ytoinf \frac{ f(y)}{h_\xi(y/\sigma)/\sigma}= \kappa \cdot b(1). 
\end{equation}
This implies \eqref{eq:upper}.

We also highlight that following identifiably issue. The two distributions 
$EGPD(\sigma,\kappa,\xi, B)$ and $EGPD(\tilde{\sigma},\kappa,\xi, \tilde{B})$ are equal if $B(0)=\tilde B(0)$ and
$$\tilde{B}(u^\kappa)= B\left( s^\kappa(u;\nu, \xi)  \right), 
$$   with $ \nu = {\tilde{\sigma}}/{\sigma}$ and 
    $s(u; \nu, \xi)= 
    H_\xi\left( \nu H_\xi^{-1}(u)\right)$.
    Therefore, $B$ and $\sigma$ cannot be let completely free in practice.

\section{Proof of Proposition \ref{prop: T(X) EGPD}}

To simplify expressions, let us assume that $\sigma=1$; the general case $\sigma \neq 1$ can easily be deduced from this particular case.
Let $F_T$ denote the cdf of $T$. It is always possible to define the function 
$$
B_T(u):=F_T\left( H^{-1}_{\xi}\left( u^{1/\kappa_T} \right) \right)
$$ for $\kappa_T>0$.
Note that 
$
B_T(0)=F_T\left( 0 \right) = \Pr(T({\bY})=0).
$

As a composition of non-decreasing functions, $B_T(u)$ is also non-decreasing for $u$ on $[0,1]$ and, by definition, we have
$F_T(y)= B_T\left(H_{\xi}(y)^{\kappa_T}\right)$.  

According to Definition~\ref{def: MGPD}, we need to show that the quantities 
$$
b_T(0^+) := \lim_{u \rightarrow 0^+} \frac{B_T(u)-B_T(0)}{u} \mbox{ and }
  b_T(1) := \lim_{u \rightarrow 1^+} \frac{B_T(u)-B_T(1)}{u-1}
$$  
are finite and non-null. 
By changing $u$ into $y=H^{-1}_{\xi}\left( u^{1/\kappa_T}\right)$ and using the Taylor expansion   
$$\Pr(Y_1> y)=1-(1-\overline{H}_{\xi}(y))^{\kappa_T} \approx \kappa_T \overline{H}_{\xi}(y)$$ for large $y$, 
we can then deduce from condition \eqref{eq: tail equivalence conditions upper T}
$$
\lim_{u \rightarrow 1^-} \frac{B_T(u)-B_T(1)}{u-1} = \lim_{y \to \infty} \frac{\Pr(T >  y)}{\kappa_T \overline{H}_{\xi}(y)}=
\alpha \frac{\kappa}{\kappa_T} \lim_{y \to \infty} \frac{\Pr(Y_1 >  y)}{\kappa \overline{H}_{\xi}(y)} = 
\alpha\frac{\kappa}{\kappa_T}   b(1)= b_T(1).
$$
Setting $\kappa_T=\kappa\gamma$, we obtain the second equation in \eqref{eq:b_T}.

Concerning the lower tail, we have for $y \geq 0$
\begin{eqnarray*}
\Pr(T \leq y)&=&\Pr(T =0)+\Pr(0<T \leq y).
\end{eqnarray*}
In addition, 
we note that as $H_\xi(y)=y +o(1)$ for   $y\rightarrow 0^+$ zero and consequently, we have   $u=H_\xi(y^{\kappa_T})= y^{\kappa_T} + o(1)$ when $u$ is near zero.
Therefore,
 \begin{eqnarray*}
     \lim_{u \rightarrow 0^+} \frac{B_T(u)-B_T(0)}{u} &=& 
      \lim_{y \rightarrow  0^+} \frac{\Pr( T  \leq  y) - \Pr( T =0)}{y^{\kappa_T}},\\
      &=& \lim_{y \rightarrow  0^+} \frac{\Pr( 0< T  \leq  y)}{y^{\kappa_T}},\\
     &=&\beta  \lim_{y \rightarrow   0^+} \frac{[\Pr(0< Y_1 \leq  y)]^\gamma }{y^{\kappa_T}}, \mbox{ by \eqref{eq: tail equivalence conditions lower}},\\
     &=&\beta  \lim_{y \rightarrow   0^+} \frac{\left( b(0^+) y^{\kappa}\right)^{\gamma} }{y^{\kappa_T}}, \mbox{ as $Y_1$ follows an EGPD($1,\kappa,\xi,B)$},\\
     &=&  \beta\; b(0^+)^\gamma, \mbox{ when $\kappa_T= \gamma \; \kappa$.} 
\end{eqnarray*}

\hfill $\square$

\section{Checking \eqref{eq: tail equivalence conditions lower} in the i.i.d.~case}
\label{sec:appendixlow}
Let $\bY=(Y_1,Y_2, \dots)^\top$ 
be a sequence of positive i.i.d.~EGPD random variables,  $Y_i \sim EGPD( \sigma,\kappa,\xi,B)$ with  $\kappa>0$ and $\xi>0$. 

~

\paragraph{\textbf{Case 1}} $T(\bY)=\max(Y_1,\dots,Y_d)$ and $B(0)=0$.\\
In this case, we can easily check that $\Pr(Y_1=0)=\Pr(T(\bY)=0) =0$ and $\Pr(T(\bY) \leq y)=\Pr(Y_1 \leq y)^d$ for $y>0$. This implies that 
$$
\cfrac{\Pr(0< T(\bY)\leq y)}{\Pr(0< Y_1\leq y)^d}
= \cfrac{\Pr(T(\bY)\leq y)}{\Pr(Y_1\leq y)^d}=\cfrac{\Pr(Y_1\leq y)^d}{\Pr(Y_1\leq y)^d}= 1.
$$
It shows that \eqref{eq: tail equivalence conditions lower} is satisfied with $\gamma=d$ and $\beta=1$.

~

\paragraph{\textbf{Case 2}} $T(\bY)=\max(Y_1,\dots,Y_d)$ and $B(0)>0$.\\
For $y>0$
\begin{eqnarray*}
\frac{\Pr(0< T(\bY)\leq y)}{\Pr(0< Y_1\leq y)}
&= &\frac{\Pr(T(\bY)\leq y)-\Pr(T(\bY)=0)}{\Pr(0<Y_1\leq y)}\\
&=&\frac{\Pr(Y_1\leq y)^d-\Pr(Y_1 = 0)^d}{\Pr(0<Y_1\leq y)}\\
&= &\frac{\left(\Pr(Y_1 = 0)+\Pr(0<Y_1\leq y)\right)^d-\Pr(Y_1 = 0)^d}{\Pr(0<Y_1\leq y)}\\
 &\rightarrow & d \Pr(Y_1=0)^{d-1}, \mbox{when $y\rightarrow 0^+$}
\end{eqnarray*}
It shows that \eqref{eq: tail equivalence conditions lower} is satisfied with $\gamma=1$ and $\beta=d B(0)^{d-1}$.

~

\paragraph{\textbf{Case 3}} $T(\bY)=\sum_{i=1}^N Y_i$, with $B(0)=0$, and $N$  is an integer value random variable, independent of $\bY$, such that $\Pr(N=1) >0$. \\
For $y>0$, we have
$$
\Pr(0< T(\bY)\leq y)=\sum_{n \geq 1}^\infty \Pr\left(\sum_{i=1}^n Y_i\leq y\right) \Pr\left (N=n \right)
$$
and thus
$$
\frac{\Pr(0<T(\bY) \leq y)}{\Pr(Y_1<y)}=\Pr(N=1) + 
\sum_{n=2}^\infty \frac{\Pr\left(\sum_{i=1}^n Y_i<y\right)}{\Pr(Y_1<y)} \Pr(N=n).
$$
Using the upper bound 
$$\Pr\left(\sum_{i=1}^n Y_i \leq y\right) \leq \prod_{i=1}^n \Pr(Y_i \leq  y)=\Pr(Y_1 \leq  y)^n$$
for $n \geq 2$, it follows that 
\begin{eqnarray*}
 \sum_{n=2}^\infty \frac{\Pr\left(\sum_{i=1}^n Y_i \leq y\right)}{\Pr(Y_1 \leq  y)} \Pr(N=n)&\leq & \sum_{n=2}^\infty \Pr(Y_1 \leq  y)^{n-1} \Pr(N=n) \\
 &\leq& \Pr(Y_1 \leq  y) \sum_{n=2}^\infty \Pr(N=n),\\
 & \leq& \Pr(Y_1 \leq y)\\
 &\rightarrow & 0, \mbox{when $y\rightarrow 0^+$}.
\end{eqnarray*}
Finally, we deduce that 
\begin{equation}
\label{eq:lowerrandomsum}
\lim_{y\rightarrow 0^+}\frac{\Pr(0< T(\bY)\leq y)}{\Pr(0<Y_1 \leq y)} =\Pr(N=1).
\end{equation}
Hence \eqref{eq: tail equivalence conditions lower} holds true with $\gamma=1$ and $\beta=\Pr(N=1)$.
\hfill $\square$

~

\paragraph{\textbf{Case 4}} Assume that $T(\bY)=\sum_{i=1}^d Y_i$ with $\Pr(Y_i=0)=B(0)>0$. \\
Let $\bY^+=(Y^+_1,Y^+_2, \dots)^\top$ 
denote an i.i.d. sequence such that $Y_i^+ \sim EGPD( \sigma,\kappa,\xi,B^+)$ with 
$$B^+(u)=\frac{B(u)-B(0)}{1-B(0)}$$ (corresponding to the distribution of $Y_i$ truncated on $(0,+\infty)$). Let $\bO=(O_1,O_2, \dots)^\top$ be an i.i.d. sequence of Bernoulli random variable independent of $\bY$ with $\Pr(O_i=0)=B(0)$. Then it can be checked that $Y_i \overset{\mathcal{D}}{=} O_i Y^+_i $ and then
$$
T(\bY) \overset{\mathcal{D}}{=}  \sum_{i=1}^d O_i Y^+_i\overset{\mathcal{D}}{=}  \sum_{i=1}^N Y^+_i$$
with $N$ the random number of non-null components in $(O_1,\dots,O_d)$. $N$ is a binomial random variables independent of  $\bY^+$ and thus \eqref{eq:lowerrandomsum} applies  
$$\lim_{y\rightarrow 0^+}\frac{\Pr(0< T(\bY)\leq y)}{\Pr(0<Y^+_1 \leq y)} =\Pr(N=1).$$
Using $\Pr(0<Y^+_1 \leq y)={\Pr(0<Y_1 \leq y)}/({1-B(0)})$ and $\Pr(N=1)=d B(0)(1-B(0))^{d-1}$ we finally deduce that 
 \eqref{eq: tail equivalence conditions lower} holds true with $\gamma=1$ and $\beta=d B(0) (1-B(0))^{d-2}$.
\hfill $\square$

~

\paragraph{\textbf{Case 5}} Assume that $T(\bY)=\sum_{i=1}^d Y_i$ with $\Pr(Y_i=0)=B(0)=0$. 

\vskip 0.5 cm
The proof is based on the following lemma which characterizes the lower tail of the sum of two independent EGPD.
\begin{lemma}\label{lem: alpha} 
Let $X_1$ and $X_2$ be two independent continuous positive random variables with pdf $f_{X_i}$. Assume  that for $i \in \{1,2\}$
$$\lim_{x\rightarrow 0^+}\frac{f_{X_i}(x)}{x^{\kappa_i-1}} =K_i$$ with $K_i>0$ and $\kappa_i>0$. Then 
$$ 
\lim_{x\rightarrow 0^+}\frac{f_{X_1+X_2}(x)}{x^{\kappa_1+\kappa_2-1}} =K_1K_2 \cB(\kappa_1,\kappa_2) $$ 
where $f_{X_1+X_2}$ denotes the pdf of $X_1+X_2$ and $\cB(a,b)$ is the beta function
$\cB(a,b)=\int_{0}^1 v^{a-1} (1-v) ^{b-1} dv$.
\end{lemma}

\textit{Proof of Lemma~\ref{lem: alpha}}. 
We need to study the limit in $0^+$ of the ratio
\begin{eqnarray*}
\frac{f_{X_1+X_2}(x)}{x^{\kappa_1+\kappa_2-1}}&=&\frac{1}{x^{\kappa_1+\kappa_2-1}}\int_0^x f_{X_1}(u)f_{X_2}(x-u)du\\
&=&\frac{1}{x^{\kappa_1+\kappa_2-2}}\int_0^1 f_{X_1}(xv) f_{X_2}(x(1-v))dv  \mbox{ with $u=vx$,}\\
& = & \int_0^1 \frac{f_{X_1}(xv)}{(xv)^{\kappa_1-1}}\frac{f_{X_2}(x(1-v))}{(x(1-v))^{\kappa_2-1}} v^{\kappa_1-1} (1-v) ^{\kappa_2-1} dv ,
\end{eqnarray*}
The dominated  convergence is then applied to obtain 
\begin{eqnarray*}
    \lim_{x\rightarrow 0^+} \frac{f_{X_1+X_2}(x)}{x^{\kappa_1+\kappa_2-1}} 
    &=& \int_{0}^1 \lim_{x\rightarrow 0^+} \frac{f_{X_1}(xv)}{(xv)^{\kappa_1-1}}
    \lim_{x\rightarrow 0^+} \frac{f_{X_2}(x(1-v))}{(x(1-v))^{\kappa_2-1}} v^{\kappa_1-1} (1-v) ^{\kappa_2-1} dv \\
    &=&   K_1 K_2 \int_{0}^1 v^{\kappa_1-1} (1-v) ^{\kappa_2-1} dv\\
    &=&   K_1 K_2  \cB(\kappa_1,\kappa_2) 
\end{eqnarray*}
\hfill $\square$

\noindent According to \eqref{eq: b(0)}, the pdf $f$ of $Y_i$ satisfies
$$
   \lim_{y\rightarrow 0^+}\frac{f(y)}{ y^{\kappa-1}}= \kappa \frac{b(0^+)}{\sigma^\kappa}. 
$$
Then using Lemma~\ref{lem: alpha} and reasoning  by recurrence, we deduce that the pdf $f_d$ of $T(\bY)=\sum_{i=1}^d Y_i$ is such that
$$
   \lim_{y\rightarrow 0^+}\frac{f_d(y)}{ y^{d\kappa-1}}=  \frac{b(0^+)^d}{\sigma^{\kappa d}} \frac{\Gamma\left(\kappa+1\right)^d}{\Gamma\left(d \kappa\right)}. 
$$
Using L'H\^opital rule, we deduce that 
\begin{eqnarray*}
\lim_{y\rightarrow 0^+}\frac{\Pr(Y_1+\dots+Y_d \leq y )}{ y^{d\kappa}}&=&\lim_{y\rightarrow 0^+}\frac{f_d(y)}{ (d\kappa) y^{d\kappa-1}}\\
&=&\frac{b(0^+)^d}{\sigma^{\kappa d}}\frac{\Gamma\left(\kappa+1\right)^d}{\Gamma\left(d \kappa+1\right)}.
\end{eqnarray*}
Using again \eqref{eq: b(0)}, we obtain the following result
\begin{eqnarray*}
\lim_{y\rightarrow 0^+}\frac{\Pr(Y_1+\dots+Y_d \leq y )}{ \Pr(Y_1<y)^d} 
& = & \lim_{y\rightarrow 0^+}\frac{\Pr(Y_1+\dots+Y_d \leq y )}{ y^{d\kappa}}\frac{y^{d\kappa}}{\Pr(Y_1<y)^d}\\
&=& \frac{\Gamma\left(\kappa+1\right)^d}{\Gamma\left(d \kappa+1\right)}.
\end{eqnarray*}

This proves that \eqref{eq: tail equivalence conditions lower} is satisfied with $\gamma=d$ and $\beta={\Gamma\left(\kappa+1\right)^d}/{\Gamma\left(d \kappa+1\right)}$.

\section{Proof of Proposition \ref{lem: CEGPDv2}}

Let $T=\sum_{i=1}^{N} Y^*_i$
 be a compound Poisson-EGPD random variable with the notations and assumptions of Definition~\ref{def:CEGPD}. Then $T$ satisfies the conditions of Proposition \eqref{prop: T(X) EGPD}; see \cite{breiman:1965} for the upper tail condition \ref{eq: tail equivalence conditions upper T}, which holds true with $\alpha=E[N]=\lambda$, and \eqref{eq:lowerrandomsum} for the lower tail condition \eqref{eq: tail equivalence conditions lower}, which holds true with $\gamma=1$ and $\beta=\Pr(N=1)=\lambda \exp(-\lambda)$. This implies that $T \sim EGPD( \sigma,\kappa,\xi, B_{\lambda})$ with $
B_{\lambda}(0)=\exp(-\lambda) $, $ b_{\lambda}(0^+)=\lambda  \exp(-\lambda)  b(0^+)$ and  $b_{\lambda}(1)=\lambda b(1)$.

Conversely, let $T \sim EGPD( \sigma,\kappa,\xi, B_T)$ be an infinitely divisible random variable. Then according to \cite[Theorem 3.2]{steutel2003infinite}, $T$ can be written as a compound Poisson distribution, $T=\sum_{i=1}^N Y_i^*$, 
with $N$ a Poisson random variable and $\bY^*=(Y_1^*,Y_2^*,...)$ an i.i.d. sequence of positive random variable independent of $N$. It remains to prove that $Y_i^*$ is EGPD.
According to \cite[Lemma 3.7.]{jessen:mikosch},    $$ \lim_{y \to \infty } \frac{\Pr( T > y) }{\Pr( Y_1^*> y) }=E[N]$$
and \eqref{eq:lowerrandomsum} implies that the lower tail of $T$ and $Y_i^*$ are also equivalent 
$$\lim_{y\rightarrow 0^+}\frac{\Pr(0< T\leq y)} {\Pr(0<Y_1^* \leq y)} =\Pr(N=1) \exp(-\lambda).$$ Using Proposition~\ref{prop: T(X) EGPD} we deduce that $Y_i^* \sim EGPD( \sigma,\kappa,\xi, B)$ (remark that the role of $T$ and $Y_i$ are symmetric in Proposition~\ref{prop: T(X) EGPD}).
\hfill $\square$

\section{Proof of Proposition~\ref{prop:monotony}}
Let $Y_1,Y_2,\dots$ be a  non-negative positive random and let $u>0$.
Remark that
\begin{equation}
\label{eq:Y1}
\Pr(Y_1 \geq u)=\Pr(A_1 \geq  u)\Pr(Y_1>0)
\end{equation}
where $A_1=Y_1^+$ denotes the positive part of $Y_1$. Under assumption \eqref{eq:mod} we have
$$A_1  \overset{\mathcal{D}}{=} \sigma_1 \sum_{i=1}^{N_1^+} X_i$$
where  $\bX=(X_1,X_2, \dots)^\top$ 
denotes a sequence of positive i.i.d. random variables with $X_i \sim EGPD( 1,\kappa,\xi,B)$ and $B(u)=u$ and $N_1^+$ the positive part of a Poisson random variable $N_1$ with mean $\lambda_1$ independent of $\bX$.

Similarly, we have
\begin{equation}
\label{eq:Yd}
\Pr(Y_d \geq u)=\Pr(A_d \geq  u)\Pr(Y_1+\dots+Y_d>0)
\end{equation}
with
$$A_d \overset{\mathcal{D}}{=}\sigma_d \sum_{i=1}^{N_d^+} X_i$$
where $N_d$ is a Poisson random variable  with mean $\lambda_d$ independent of $\bX$. 

Under the assumptions of Proposition~\ref{prop:monotony}, we have $\lambda_1 \leq \lambda_d$ and $\sigma_1 \leq \sigma_d$. It implies that $N_1$ is smaller than $N_d$ in the likelihood ratio order, i.e. the ratio ${\Pr(N_d=n)}/{\Pr(N_1=n)}$ is increasing in  $n$. Using \cite[Theorem~1.C.6.]{shaked2007stochastic} we deduce that  $N_1^+$ is smaller than $N_d^+$ in the likelihood ratio order and thus also in the usual stochastic order. Then \cite[Theorem~1.A.4]{shaked2007stochastic} implies that $A_1$ is  smaller than $A_d$ in the usual stochastic order, meaning that for any $u>0$
\begin{equation}
\label{eq:A1Ad}
\Pr(A_1 \geq u) \leq \Pr(A_d \geq u).
\end{equation}
Combining the inequality $$\Pr(Y_1>0) \leq  \Pr(Y_1+\dots+Y_d>0)$$
with equations \eqref{eq:Y1}, \eqref{eq:Yd}, \eqref{eq:A1Ad}, we  deduce that \eqref{eq: constraint} holds true.

\section{Some properties of $EGPD(\sigma,\kappa,\xi, \lambda)$}
\label{sec:Panjer}

Let $\bY^*=(Y^*_1,Y^*_2, \dots)^\top$ 
be a sequence of  i.i.d.~EGPD random variables,  $Y^*_i \sim EGPD( \sigma,\kappa,\xi,B)$, with $B(u)=u$,  $\kappa>0$ and $\xi>0$.  Let $N$ be  a Poisson random variable  with mean $\lambda$ independent of $\bY^*$. We denote
$    Y= \sum_{i=1}^{N} Y^*_i$
and $A=Y^+$ the positive part of this distribution. According to Definition~\ref{def:CEGPD+} we have $A \sim EGPD(\sigma,\kappa,\xi, \lambda)$.

Remark that  for $y>0$, 
\begin{eqnarray*}
P(A \leq y) &= & P(Y \leq y|Y>0)\\
&= & \frac{P(Y \leq y}{P(Y>0)},
\end{eqnarray*}
and thus
\begin{equation}
\label{eq : linkAY}
P(A \leq y) = \frac{P(Y \leq y)}{1-exp(-\lambda)}. 
\end{equation}

Properties of $A$, such as its first two moments for example, can then be deduced from the general properties of compound Poisson distributions \cite[see, e.g.][]{ross2014introduction}. The pdf of $A$  is of particular interest for this work since it appears in the definition of composite likelihood function \ref{eq:CL}. Several methods have been proposed in the literature to compute numerical approximations of the pdf of compound Poisson distributions \cite{embrechts2009panjer}. In this work, we use the Panjer recursions \cite{panjer1981recursive}, which takes a discrete distribution as input. 

We thus first replace $Y^*_i$  with  its  discrete version $E^*_i(h)$ concentrated on $\{h,2h,\ldots,\}$ where $h=0.2$ is the built-in precision of existing precipitation gauges in Section~\ref{sec: inference and application}.
Using the general expression \eqref{eq: F bar} of the cdf of the EGPD distribution with $B(u)=u$, we obtain the  probability  function of $E^*_i(h)$ 
\begin{equation}
f_j =\Pr(E^*_i(h)=jh)=H_\xi^\kappa(\frac{jh} \sigma) - H_\xi^\kappa(\frac{(j-1)h}{\sigma})
\end{equation}
for $j \in \{1,2,\dots\}$.
  For compound Poisson  distribution, the recursive Panjer formula for $p_a=\Pr(\sum_{i=1}^{N_d}  E^*_i (h)=ah)$, $a \in \{0,2,\dots\}$, reduces to
\begin{equation}\label{eq:panjer}
	p_a=\left\{
	\begin{array}{lc}
		\exp(-\lambda)     &\mbox{if } a=0 \\
		(\lambda/{a})\sum_{j=1}^a j f_jp_{a-j}    &\mbox{if } a\ge 1 
	\end{array}
	\right.
\end{equation}

\section{Simulation results}
\label{sec:Appendixsimu} 

In this Appendix, synthetic rainfall data at the finer time scale is simulated as an i.i.d. sequence of a compound Poisson-EGPD, see Definition~\ref{def:CEGPD} with $B(u)=u$. In this idealized setting, it can be checked that the positive aggregated data defined by \eqref{eq:amount} satisfy \eqref{eq:mod}  and \eqref{eq:modpara} where  $\sigma_d=\sigma$ and $\lambda_d=d \lambda$ are log-polynomials of order $p=0$ and $q=1$.  To mimic our application setup,  the parameters are fixed to $\kappa=.3$, $\xi =.25$, $\sigma=1$, $\lambda=.01$, the sample size is equivalent to $36$ months of $6$ minute rainfall data and we let $p=q=3$ in the estimation procedure described.

\begin{figure}
    \centering
    \includegraphics[width=13cm]{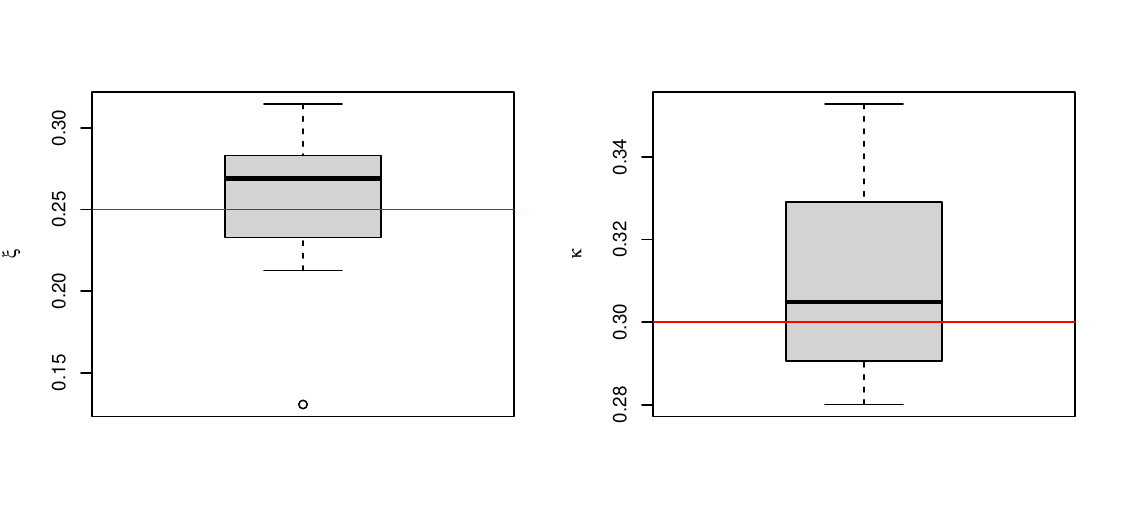}
    \includegraphics[width=6.5cm]{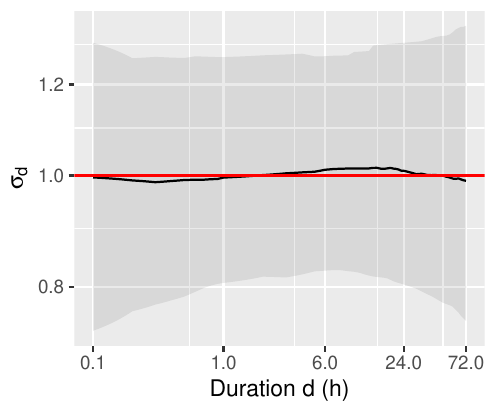}
    \includegraphics[width=6.5cm]{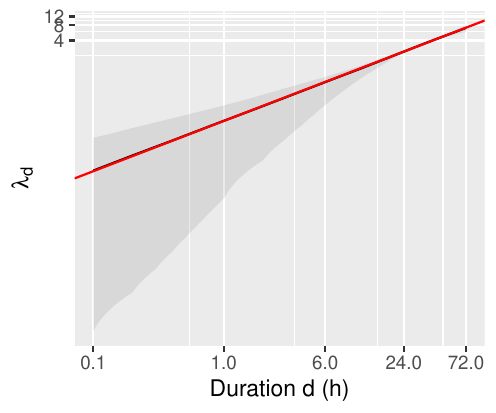}

    \caption{Empirical distribution of the parameter estimates obtained using the simulation setup described in \ref{sec:Appendixsimu}. The red lines correspond to the truth. On the bottom plots, the black lines correspond to the median, grey areas to $95\%$ fluctuations interval computed by repeating the experiment $500$ time.}
    \label{fig:simu}
\end{figure}

Figure \ref{fig:simu} displays our estimates of $\xi$, $\kappa$, $\sigma_d$ and $\lambda_d$ 
in the top left, top right, bottom left and bottom right panels, respectively. 
Overall, the parameters are well estimated. 

\newpage
\bibliographystyle{elsarticle-num} 
\bibliography{bibfilenew}

\end{document}